\documentclass[10pt,letterpaper,twocolumn]{article} 

\usepackage{ol2}
\usepackage{amsmath,amssymb}
\usepackage[colorlinks=true,breaklinks=true,hypertex]{hyperref}
\usepackage{dcolumn}

\usepackage{epsfig}
\usepackage{graphicx}

\newcommand{\kp}{k_1}

\newcommand{\mic}{~\rm \mu m}

\begin{document}

\twocolumn[
\title{Higher-order Kerr effect and harmonic cascading in gases}

\author{Morten Bache$^{1*}$, Falk Eilenberger$^2$ and Stefano Minardi$^{2}$}

\address{
$^1$DTU Fotonik, Department of Photonics Engineering,
Technical University of Denmark,
  DK-2800 Kgs. Lyngby, Denmark.\\
$^2$Institute of Applied Physics, Abbe Center of Photonics, Friedrich-Schiller-Universit\"at Jena, 
07743 Jena, Germany\\
$^*$Corresponding author: moba@fotonik.dtu.dk
}

\begin{abstract}

The higher-order Kerr effect (HOKE) has been recently advocated to explain measurements of the saturation of the nonlinear refractive index in gases. Here we show that cascaded third-harmonic generation results in an effective fifth order nonlinearity that is negative and significant. Higher-order harmonic cascading will also occur from the HOKE, and the cascading contributions may significantly modify the observed nonlinear index change. At lower wavelengths cascading increases the HOKE saturation intensity, while for longer wavelengths cascading will decrease the HOKE saturation intensity.
\end{abstract}


]
In homogeneous, isotropic media the nonlinear variation of the refractive index can be expressed in terms of a power series of the light intensity $I$: $\Delta n=n_2 I+n_4 I^2 +...+n_{2m}I^m$. However, higher-order nonlinearities have usually been regarded as negligible, as compared to the leading third-order nonlinearity $n_2$ (Kerr effect). A recent experiment aimed at characterizing the higher-order Kerr effect (HOKE) \cite{Loriot:2009} has sparked a vivid debate on the actual role of these higher-order terms in nonlinear beam-propagation dynamics. Indeed, the effective saturation of the HOKE nonlinearity could lead to filamentation of femtosecond pulses in gases without the plasma playing an active role \cite{bejot:2010}. The mechanism advocated in this case to arrest collapse of the wave packet was the saturation of the nonlinearity rather than the formation of plasma, as commonly accepted \cite{Couairon:2007}.
Experiments have either confirmed \cite{bejot:2011a,Petrarca:2012,Kartashov:2012b} or ruled out \cite{Polynkin:2011,Ariunbold:2012,Wahlstrand:2011}
a preponderant role of HOKE in filamentation dynamics. 

In the strong cascading limit cascading of harmonics gives effective higher-order nonlinearities that compete with the inherent material ones. 2nd-order $\chi^{(2)}:\chi^{(2)}$ cascading gives an effective $\chi^{(3)}$ nonlinearity \cite{Ostrovskii:1967,stegeman:1996}, which can be made ultrafast and negative (self-defocusing)
and even strong enough to overcome the self-focusing
material Kerr nonlinearity \cite{zhou:2012}. Similarly, 3rd-order $\chi^{(3)}:\chi^{(3)}$ cascading gives an effective $\chi^{(5)}$ \cite{Saltiel:1997}. Here we study cascading contributions to the nonlinear HOKE coefficients in gases. This is unlike previous investigations that focussed on harmonic yield from cascading \cite{Polynkin:2011,Ariunbold:2012,Kartashov:2012b}. Cascading relies on a phase mismatch $\Delta k_{1m}=k(\omega_m)-mk(\omega_1)\neq 0$ between the fundamental wave (FW, $\omega_1$) and the $m$th harmonic ($\omega_m$). After two coherence lengths $2\pi/|\Delta k_{1m}|$ back-conversion to the FW is complete, and as $\Delta k_{1m}\neq 0$ the harmonic has a different phase velocity than the FW. Thus, the back-converted FW photons are phase shifted with respect to the unconverted FW photons. For a strong phase mismatch $\Delta k_{1m}L\gg 1$ ($L$ is the interaction length), this process repeats many times and the FW feels a nonlinear phase shift parameterized by an \textit{effective} higher-order nonlinear index $n_{2m,\rm casc}\propto -[\chi^{(m)}]^2/\Delta k_{1m}$. We show that in gases cascading can affect the nonlinear index change and in particular the saturation intensity. Specifically, to lowest order cascading comes from a phase-mismatched third-harmonic generation (THG) process, which we show generates an $n_4$ term that is negative and of significant strength. Similarly, cascading of HOKE terms gives contributions to subsequent orders of nonlinearity ($n_6$, $n_8,\dots$) with values comparable with recent experimental \cite{Loriot:2009} and theoretical HOKE coefficients \cite{bree:2011}. This implies that in the strong cascading limit, defined later, cascading may contribute significantly to the nonlinear dynamics.

\begin{table*}
  \centering
  \caption{
  Material \cite{Loriot:2009} and cascading nonlinearities at $\lambda_1=0.8\mic$, $p=1$ atm and $T=293$ K.
  }
\label{tab:disp2}
  \begin{tabular}{l|c c|c c|c c|c c|c c}
    & ${\Delta k_{13}}^a$ &	$n_2$ &	$n_4$ &	$n_{4,\rm casc}$	& $n_6$	 & $n_{6,\rm casc}$ & $n_8$ & 	 $n_{8,\rm casc}$ &	$n_{10}$ &	$n_{10,\rm casc}$ \\
    \hline
Ar	    & 442 & 1.0E-23 & -3.6E-42 & -1.8E-42 & 40E-59 & 0.2E-59 & -17.2E-76 & -3.0E-76 & 0.9E-93 & 2.0E-93 \\
Air	    & 482 & 1.2E-23 & -15.0E-42  & -2.3E-42 & 21E-59 & 0.8E-59 & -8.0E-76 & -1.8E-76	 & -	 & 1.0E-93 \\
O$_2$	& 772 & 1.6E-23 & -51.6E-42  & -2.6E-42 & 48E-59 & 2.2E-59 & -21.0E-76 & -3.3E-76 & -	 & 2.2E-93 \\
N$_2$	& 501 & 1.1E-23 & -5.3E-42 & -1.9E-42 & 14E-59 & 0.3E-59 & -4.4E-76 & -1.0E-76 & -	 & 0.5E-93 \\
  \end{tabular}\\
  {\footnotesize $^a$Calculated using Sellmeier equations \cite{Borzsonyi:2008,Zhang:2008}. All units are SI (i.e. $[\Delta k_{13}]=\rm m^{-1}$, $[n_2]=\rm m^2/W$ etc.). Note: the cascading calculations gave orders beyond $n_{10,\rm casc}$, which are included in Fig. \ref{fig:Dn}. The propagated error in the leading cascading term $n_{4, \rm casc}$ is 12-35\%.}
\end{table*}

We now discuss cascaded THG,
and will generalize to higher harmonics later. In the slowly-varying envelope approximation the mks electric-field coupled-wave propagation equations for THG $\omega_1+\omega_1+\omega_1\rightarrow \omega_3$ in the plane-wave limit in an isotropic lossless medium are
\begin{align}\label{eq:FW}
\left(  i
\partial_z-\tfrac{1}{2}k_2(\omega_1)\partial_{\tau\tau}\right ) E_1
\nonumber\\
+\Gamma_{\omega_1}^{(3)}\left[E_1|E_1|^2 +  2E_1|E_3|^2 + (E_1^*)^2 E_3 e^{i\Delta k_{13} z}
\right] &=0
\\
\left(  i
\partial_z-id_{13}\partial_\tau
-\tfrac{1}{2}k_2(\omega_3)\partial_{\tau\tau}\right ) E_3
\nonumber\\
+\Gamma_{\omega_3}^{(3)}\left[E_3|E_3|^2
+
  2E_3|E_1|^2 + \tfrac{1}{3}E_1^3 e^{-i\Delta k_{13} z}
\right] &=0
\label{eq:TH}
\end{align}
$k(\omega)=n({\omega})\omega/c$ is the wave number, $n(\omega)$ the linear refractive index, $k_m(\omega)\equiv d^m k(\omega)/d\omega^m$ the dispersion coefficients, $d_{13}=\kp(\omega_1)-\kp(\omega_3)$ the group-velocity mismatch, 
and $\Gamma_{\omega_j}^{(3)}= 3\omega_j\chi^{(3)} /8cn({\omega_j})$,  $j=1,3$. 
In the strong cascading limit, $\Delta k_{13}L\gg 1$ and no FW depletion are assumed \cite{bache:2007a}, so third harmonic (TH) self- and cross-phase modulation in Eq. (\ref{eq:TH}) are negligible and only the $E_1^3/3$ term is relevant. The ansatz $E_{3,\rm casc}(z,\tau)=e^{-i\Delta k_{13} z}\psi_3(\tau)$ makes Eq. (\ref{eq:TH}) a temporal ODE. Neglecting dispersion for the moment gives $E_{3,\rm casc}(z,\tau)=-e^{-i\Delta k_{13} z}E_1^3(z,\tau)\Gamma_{\omega_3}^{(3)}/(3\Delta k_{13})$: the TH is effectively slaved to the FW.
Inserting this result in Eq. (\ref{eq:FW}) yields the nonlinear term $\Gamma_{\omega_1}^{(3)}E_1|E_1|^2
\left[1 -|E_1|^2 \Gamma_{\omega_1}^{(3)}/\Delta k_{13}+  2|E_1|^4({\Gamma_{\omega_1}^{(3)}}/\Delta k_{13})^2 \right]$, where we used $\Gamma_{\omega_3}^{(3)}/3\simeq \Gamma_{\omega_1}^{(3)}$ and
$n({\omega_1})/n({\omega_3})\simeq 1$.
Converting to intensity
$E_1\rightarrow [2/\varepsilon_0 n({\omega_1}) c]^{1/2}A_1$
we get
$(n_2A_1|A_1|^2 + n_{4,\rm casc}A_1|A_1|^4+ n_{6,\rm casc}^{(3)}A_1|A_1|^6)\omega_1/c$, where $n_2
=3\chi^{(3)}/4\varepsilon_0cn^2({\omega_1})$ is the Kerr nonlinearity. Thus, the cascaded TH gives rise to 5th- and 7th-order \textit{self-action} nonlinearities, where the 5th-order term is
\vspace{-0.3cm}
\begin{align}\label{eq:casc}
    n_{4,\rm casc}\equiv -\frac{n_2^2}{\Delta k_{13}}\frac{\omega_1}{c}
\end{align}
This adds to the material quintic nonlinearity $n_4$ so $n_{4,\rm tot}=n_4+n_{4,\rm casc}$. In other words, when THG is included in the description, $\chi^{(3)}:\chi^{(3)}$ cascading leads to a 5th-order HOKE term that is tunable in sign and magnitude. In the case we consider, a nonlinear medium without waveguide contributions pumped far from resonances, $\Delta k_{13}>0$, so $n_{4,\rm casc}$ is always negative. 

At the next order we get $n_{6,\rm casc}=n_{6,\rm casc}^{(3)}+n_{6,\rm casc}^{(5)}$. The former term was derived above, and 
the latter term 
comes from cascaded THG in the quintic nonlinear FW term (governed by $n_4$), i.e. $\chi^{(3)}:\chi^{(5)}$ cascading. These add to the material $n_6$, so $n_{6,\rm tot}=n_6+n_{6,\rm casc}$, and read
\begin{align}\label{eq:n6casc}
    n_{6,\rm casc}^{(3)}\equiv \frac{n_2^3}{n(\omega_3)\Delta k_{13}^2}\frac{\omega_1^2}{c^2}, \quad n_{6,\rm casc}^{(5)}\equiv -\frac{5n_2n_4}{2\Delta k_{13}}\frac{\omega_1}{c}
\end{align}

The expressions for the following orders are lengthy, but briefly in the strong cascading limit the harmonic is reduced to $E_{m,\rm casc}\propto-(\Delta k_{1m})^{-1}E_1^m e^{-i\Delta k_{1m}z}$. 
Inserting these cascading terms in the FW equation, which contains all possible harmonic self- and cross-phase modulation terms, gives the cascaded higher-order self-action nonlinear coefficients. 
In this way we calculated HOKE contributions from 3rd- to 11th-harmonic generation (nonlinearities governed by $n_2$ to $n_{10}$), and Table 1 shows the cascading terms up to $n_{10,\rm casc}$ 
for various gases at $\lambda_1=0.8\mic$.
Note firstly that the cascading terms have similar strengths as the material nonlinearities at that level. Secondly, the $\Delta k_{13}$ values for THG imply that the strong cascading limit can be achieved with propagation lengths fractions of a meter (the higher-order $\Delta k_{1m}$ are always larger than $\Delta k_{13}$, so they are less critical). 

\begin{figure}[tb]
\includegraphics[height=4.9 cm]{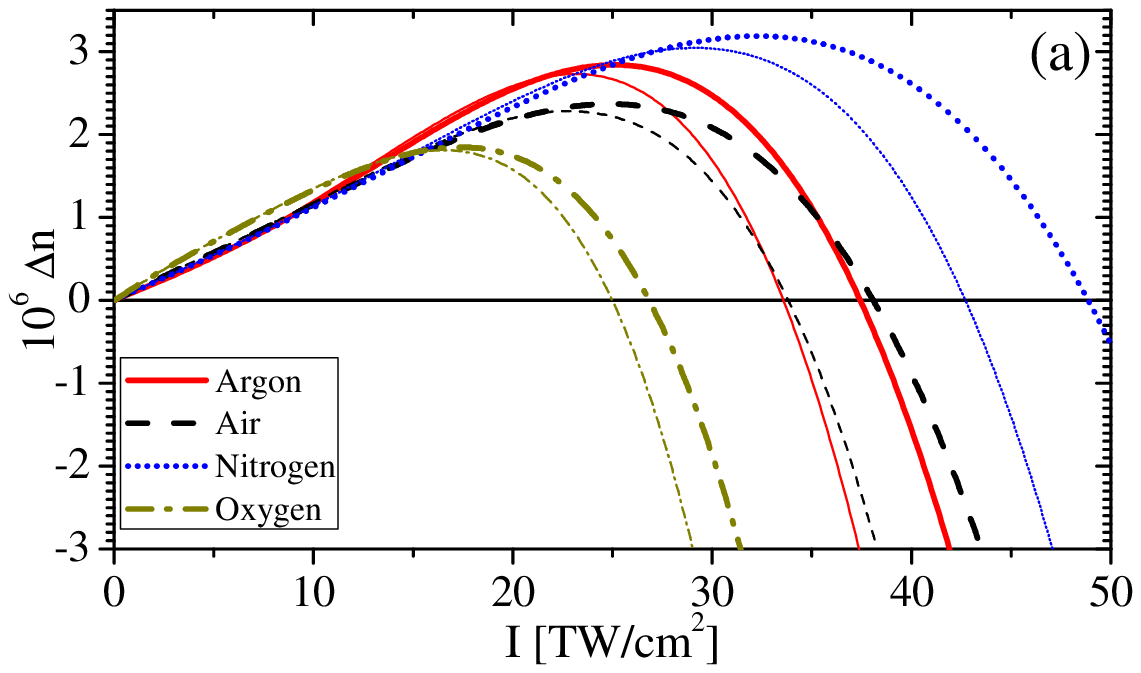}
\includegraphics[height=4.9 cm]{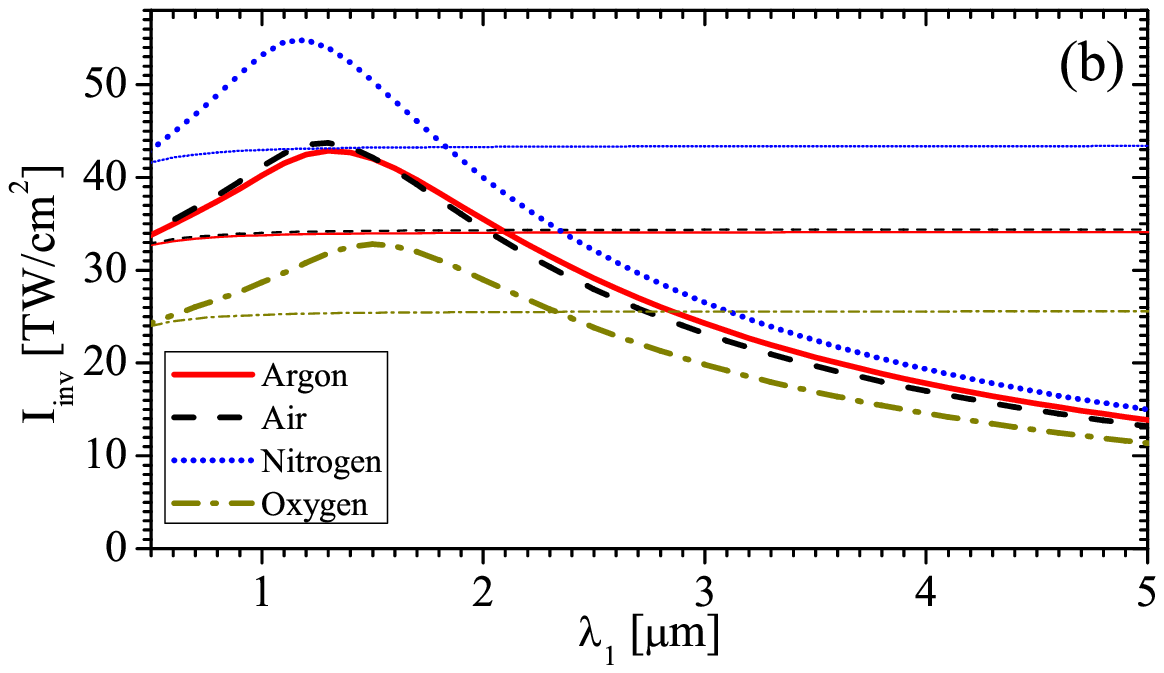}
\caption{(Color online) (a) The nonlinear index change $\Delta n$ vs. pump intensity for $\lambda_1=0.8\mic$, 
(b) 
the inversion intensity $I_{\rm inv}$
vs. FW wavelength. In both plots the thin lines are the data from \cite{Loriot:2009}, while the bold lines include cascading corrections.
In (b) the nonlinear coefficients include frequency dispersion using Miller's formula \cite{Ettoumi:2010}.
}
\label{fig:Dn}
\end{figure}


Figure \ref{fig:Dn}(a) shows the impact of cascading at $\lambda_1=0.8\mic$, where we plot the variation of the nonlinear refractive index as a function of the pump intensity for the gases in Table \ref{tab:disp2}.
It is clear that the inclusion of the cascading effect shifts the inversion intensity $I_{\rm inv}$, where $\Delta n=0$, to higher values.
However, at longer pump wavelengths the flattened gas dispersion enhances cascading. As a result the inversion intensity becomes smaller and departures from the ideal Kerr effect is observable already at low intensities. In Fig. \ref{fig:Dn}(b) the inversion intensities are plotted vs. pump wavelength to summarize these trends. Without cascading, only a slight wavelength scaling \cite{Ettoumi:2010} of the nonlinearities is seen. When including cascading the inversion intensity peaks in the near-IR, and is pushed to become lower than without cascading at the beginning of the mid-IR. The calculations are based on the HOKE material nonlinearities from \cite{Loriot:2009}, but we found similar results using the coefficients from \cite{bree:2011} (seemingly, the HOKE coefficients of \cite{Loriot:2009} and \cite{bree:2011} do not match, but this is in part due to the large difference in the number of HOKE orders). Note that the diatomic gases (air, O$_2$ and N$_2$) also have fractional contributions from rotational and vibrational Raman effects that contribute to the overall observed nonlinearity as well, and this might make it more difficult to observe effects of cascading.

The strong cascading limit \cite{bache:2007a} assumes that the coherence length $l_{\rm coh}=\pi/|\Delta k_{1m}|$ is much smaller than (a) the interaction length ($\Delta k_{1m}L \gg 1$) and (b) the characteristic nonlinear length ($\Delta k_{1m}/[\Gamma_{\omega_m}^{(m)}E_{1,\rm in}^{m-1}]\gg 1$). To lowest order no pump depletion is found: $E_{3,\rm casc}$ above Eq. (\ref{eq:casc}) is just phase modulated in $z$. When $\Delta k_{1m}L \gg 1$ cascading gives an effective Kerr-like nonlinearity ($\Delta n$ independent on $z$), and when $\Delta k_{1m}/[\Gamma_{\omega_m}^{(m)}E_{1,\rm in}^{m-1}] \gg 1$ the approximation $\Delta n=\sum_j n_{2j} I^j$ holds. For $\Delta k_{1m}L\simeq 1$ this breaks down as the cascaded harmonic becomes amplitude modulated by $z{\rm sinc}(\Delta k_{1m} z/2)$. The cascading terms must then be corrected by $1-{\rm sinc}(\Delta k_{1m} L)$ (resolving the unphysical divergences at $\Delta k_{1m}=0$). 
Nonetheless, even for $\Delta k_{1m} L\lesssim 1$, strong cascading is possible if $\Delta k_{1m}/[\Gamma_{\omega_m}^{(m)}E_{1,\rm in}^{m-1}] \lesssim  1$ (strong FW depletion), 
but the cascading will then grow step-wise in $z$ \cite{stegeman:1996}, so the cascading nonlinear coefficients cannot be parameterized. 

The increased cascading contributions predicted at longer wavelengths in Fig. \ref{fig:Dn}(b) rely on reduced phase-mismatch values. 
Thus, fulfilling $\Delta k_{13} L\gg 1$ can become problematic except for very long cells.  Filamentation experiments will therefore make an ideal test bed for cascading as they inherently use very long interaction lengths, where the sinc$(\Delta k_{13} L)$-oscillations vanish. Alternatively, a solution could be to increase the pressure: the inversion intensity remains the same, but the phase mismatch values increase and overall we find that the cascading contributions to $\Delta n$ are enhanced.

The experiment behind the HOKE coefficients \cite{Loriot:2009} used a pump-probe interaction length $L\simeq 1$ mm \cite{lavorel:2012}. Since the leading phase-mismatch values (see Table \ref{tab:disp2}) are around 1/mm or less, the cascading contribution is reduced because the $1-{\rm sinc}(\Delta k_{13} L)$ term must be applied. Specifically, we found that cascading corrections to all HOKE terms for the various gases are within the error bars of the measurements. Thus, the HOKE values in \cite{Loriot:2009} correspond to the intrinsic material nonlinearities. 


Even if cascading relies on a chain of up- and down-conversion steps, it can be considered instantaneous, much like an electronic Kerr nonlinearity: with full dispersion, the cascading term in frequency domain becomes $\tilde E_{3,\rm casc}(z,\Omega)\propto-\Delta k_{13}^{-1}{\mathcal F}[E_1^3(z,\tau)]R_{\rm casc}(\Omega)$ where ${\mathcal F}[\cdot]$ denotes the Fourier transform and 
$R_{\rm casc}(\Omega)=\Delta k_{13}/[k(\Omega+\omega_3)-k_1(\omega_1)\Omega-3 k(\omega_1)]$ is the cascading "response" function \cite{bache:2007a}.  Neglecting higher-order dispersion the criterion $d_{13}^2<2 k_2({\omega_3})\Delta k_{13}$ approximately determines when the cascading response is nonresonant and thus broadband \cite{bache:2007a}. We checked that this inequality is always fulfilled, and moreover since octave-spanning bandwidths are supported 
the corresponding temporal response of the cascading is extremely fast \cite{zhou:2012}. 

Concluding, cascading in Kerr media from phase-mismatched harmonic generation may significantly contribute to the observed nonlinear index change in gases. The calculations rely on the strong phase-mismatch limit, and the lowest order contribution from $n_2$, where $\chi^{(3)}:\chi^{(3)}$
cascading from a phase-mismatched third-harmonic gives an effective $n_4$ term, is always negative and significant 
compared to the material $n_4$. A unidirectional pulse propagation equation can model such harmonic dynamics with a proper nonlinear expansion \cite{Kolesik2006}.
For higher-order harmonics the conclusions were similar; such terms rely on including higher-order material nonlinearities. At 800 nm the cascading from $n_2$ and higher-order coefficient ($n_4$ up to $n_{10}$) \cite{Loriot:2009} delayed the saturation point to higher intensities than what the material nonlinearities dictate. This might explain why indications of saturation for near-IR filaments have been elusive. Instead for longer (mid-IR) wavelengths, cascading pushed the saturation point to lower intensities, which will enhance saturation and possibly affect filamentation.

Acknowledgements: The Danish Council for Independent Research projects  274-08-0479 and 11-106702.




%

\end{document}